\documentclass[pdftexify,conference]{IEEEtran}
\usepackage{cite}
\usepackage{amsmath, amssymb, amsfonts, bbm, amsthm, eqparbox}
\usepackage{array, float}
\usepackage{graphicx,epsfig}
\usepackage[tight,footnotesize]{subfigure}
\usepackage[colorlinks,linkcolor=black,urlcolor=black,citecolor=black]{hyperref}
\usepackage{url}
\usepackage[latin1]{inputenc}
\usepackage[T1]{fontenc}
\usepackage[ruled,  vlined, linesnumbered]{algorithm2e}
\usepackage{multirow}
\usepackage{multicol}
\usepackage{color}
\IEEEoverridecommandlockouts

\newcommand{\MYHeader}{IV Workshop-school on Quantum Computation and Information (WECIQ 2012)}

\makeatletter

\def\ps@headings{%
\def\@oddhead{\MYHeader \hfil}
\def\@evenhead{\MYHeader \hfil }
\def\@oddfoot{\ }%
\def\@evenfoot{\ }}

\def\ps@IEEEtitlepagestyle{%
\def\@oddhead{\MYHeader \hfil}
\def\@evenhead{\MYHeader \hfil}
\def\@oddfoot{\ }%
\def\@evenfoot{\ }}

\makeatother

\newtheorem{theorem}{Theorem}
\newtheorem{corollary}{Corollary}
\newtheorem{lemma}{Lemma}

\newtheorem{definition}{Definition}

\newtheorem{property}{Property}

\input{Qcircuit.tex}

\begin{document}

\title{Cayley graphs and analysis of quantum cost for reversible
  circuit synthesis}

\author{A.~C.~Ribeiro, C.~M.~H.~de~Figueiredo, F.~L.~Marquezino,
  L.~A.~B.~Kowada \thanks{This work was supported in part by PIQDTec,
    CNPq and FAPERJ. A.~C.~Ribeiro, C.~M.~H.~de~Figueiredo and
    F.~L.~Marquezino are with the Universidade Federal do Rio de
    Janeiro, Brazil. E-mail: rcandre@cos.ufrj.br, celina@cos.ufrj.br
    and franklin@cos.ufrj.br. L.~A.~B.~Kowada is with Universidade
    Federal Fluminense, Brazil. E-mail: luis@vm.uff.br. A. C. Ribeiro
    is with Instituto Federal Goiano, Brazil.  }}

\maketitle

\begin{abstract}
  We propose the theory of Cayley graphs as a framework to analyse
  gate counts and quantum costs resulting from reversible circuit
  synthesis.  Several methods have been proposed in the reversible
  logic synthesis literature by considering different libraries whose
  gates are associated to the generating sets of certain Cayley
  graphs. In a Cayley graph, the distance between two vertices
  corresponds to the optimal circuit size. The lower bound for the
  diameter of Cayley graphs is also a lower bound for the worst case
  for any algorithm that uses the corresponding gate library. In this
  paper, we study two Cayley graphs on the Symmetric Group $S_{2^n}$:
  the first, denoted by $I_n$, is defined by a generating set
  associated to generalized Toffoli gates; and the second, the
  hypercube Cayley graph $H_n$, is defined by a generating set
  associated to multiple-control Toffoli gates. Those two Cayley
  graphs have degree $n2^{n-1}$ and order $2^n!$. Maslov, Dueck and
  Miller proposed a reversible circuit synthesis that we model by the
  Cayley graph $I_n$. We propose a synthesis algorithm based on the
  Cayley graph $H_n$ with upper bound of $(n-1)2^{n}+1$
  multiple-control Toffoli gates. In addition, the diameter of the
  Cayley graph $H_n$ gives a lower bound of $n2^{n-1}$.
\end{abstract}


\section{Introduction}

An important feature of the circuit model of quantum computation is reversibility.  This isa consequence of the evolution postulate of quantum mechanics, which
states that the time-evolution of the state of a closed quantum system
is described by a unitary operator~\cite{nielsen00}. Therefore, the
theory of reversible computation is one of the foundations of quantum
computation. In any reversible circuit---classical or quantum---the
output contains sufficient information to reconstruct the input, i.e.,
no input information is erased~\cite{toffoli}. This aspect of
reversible computation has important physical consequences. For
instance, it is well established that conventional logic gates lead to
at least $kT \ln 2 $ energy dissipation per irreversible bit
operation, where $k$ is Boltzmann's constant and $T$ is the absolute
temperature of the circuit~\cite{rolf}. Therefore, logical circuits
with almost zero power dissipation will only be possible if they are
built from reversible gates~\cite{ben}. Thus, reversible computers can
be economically more interesting for low-power design than a computer
with conventional circuits. The field of reversible computing also
draws motivation from several sources, such as signal processing,
cryptography, computer graphics, nano and photonic circuits, just to
mention a few~\cite{survey}.

A set of reversible gates is needed to design reversible
circuits. Group Theory has recently been employed as a tool to analyse
reversible logic gates and investigate generators for the group of
reversible gates~\cite{leo,alexis}. In this work, we study Cayley
graphs associated to the Symmetric Group $S_{2^n}$ in order to analyse
reversible circuit synthesis methods. Each method has been proposed by
considering different libraries whose gates are associated to the
generating sets of certain Cayley graphs.

Several properties of these Cayley graphs---such as degree, distance
and diameter---are considered.  The degree of the Cayley graph is
exactly the size of the generating set, which in turn corresponds to
the size of the gate library. The distance between two vertices
corresponds to the optimal circuit size---each gate produces an edge
of the Cayley graph, so the circuit size corresponds to the distance.
Finally, an important property of Cayley graphs is that the lower
bound for their diameter is also a lower bound for the worst case of
any algorithm that uses the corresponding gate library.

Our goal is to analyse reversible circuit synthesis based on Cayley
graphs. We present two Cayley graphs on the Symmetric Group $S_{2^n}$:
the first, denoted by $I_n$, is defined by a generating set associated
to generalized Toffoli gates (G-Toffoli); the second, the hypercube
Cayley graph $H_n$, is defined by a generating set associated to
multiple-control Toffoli gates (MC-Toffoli). Maslov, Dueck and
Miller~\cite{mar,mar1} proposed a reversible circuit synthesis that we
model by the Cayley graph $I_n$. We propose a synthesis algorithm
based on the Cayley graph $H_n$ with upper bound of $(n-1)2^{n}+1$
multiple-control Toffoli gates. In addition, the diameter of the
Cayley graph $H_n$ gives a lower bound of $n2^{n-1}$.

This paper is organized as follows. In Section 2, we introduce
notation and review some basic concepts on circuits and group theory
that will be necessary throughout the paper. We also describe two
algorithms for circuit synthesis: the first, based on G-Toffoli gates;
and the second, based on MC-Toffoli gates. In Section 3, we present
the analysis of the circuit synthesis based on G-Toffoli gates, which
we model by Cayley graph $I_n$. In Section 4, we present the analysis
of the circuit synthesis based on MC-Toffoli gates, which we model by
Cayley graph $H_n$. In Section 5, we present our conclusions.

\section{Preliminaries}

\subsection{Graph Theory and Cayley graphs}

Let $(G,\cdot)$ be a finite group with identity element denoted by
$\iota$. A subset $C$ of this group is a \emph{generating set} if
every element of $G$ can be expressed as a finite product of elements
in $C$. We also say that $G$ is generated by $C$.

\begin{definition}
  Let $C$ be a generating set for a group $G$. We say that a directed
  graph $\Gamma(V,E)$ is a \emph{Cayley graph} associated to a group
  with generating set $(G,C)$, if there exists a bijection mapping
  every vertex $v \in V$ to a group element $g \in G$, such that group
  elements $(g,h) \in G$ are connected by a directed edge $(g,h) \in
  E$ if and only if exists $c\in C$ such that $h=c \cdot g$.
\end{definition}

If $\iota \notin C$, then there are no loops  (i.e., edge between a same element) in $\Gamma$, which we define as the identity free property. If $c \in C$ implies $c^{-1} \in
C$, then for every edge from $g$ to $g \cdot c$, there is also an edge
from $g \cdot c$ to $(g \cdot c) \cdot c^{-1}=g$, which we define as
the symmetry condition. The Cayley graph with identity free property
and symmetry condition is an undirected graph. In this paper, we only
consider undirected graphs.

Let $A$ be a finite set and $f : A \rightarrow A$ a bijective
function, i.e., a permutation. For example, $\pi=[1 \ 3 \ 2 \ 0]$ is a
permutation over $\{ 0,1,2,3 \}$ where $\pi[0]=1$, $\pi[1]=3$,
$\pi[2]=2$ and $\pi[3]=0$. The set of all $n!$ permutations on
$A=\{0,1,\cdots,{n-1}\}$ with function composition operation forms the
Symmetric Group $S_n$ on $A$.

\begin{definition}
  The \emph{distance} $d(u,v)$ between the vertices $u$ and $v$ in a
  graph is the number of edges in a shortest path connecting them.
\end{definition}

\begin{definition}
  The \emph{diameter} $D$ is the largest distance among all pairs of
  vertices.
\end{definition}

In Cayley graphs, the problem of finding the distance among all pairs
of vertices, is equivalent to finding the minimum length sequence that
creates the element $p$ from $\iota$, see~\cite{vadapalli}. So, in
order to find the diameter, it is sufficient to calculate the greatest
distance between the identity vertex and all other vertices.

\begin{definition}
  Let $\pi_b$ and $\sigma_b$ be the binary representations of
  permutations $\pi$ and $\sigma$, respectively. \emph{Hamming
    distance} $d_H(\pi,\sigma)$ is the number of positions in $\pi_b$
  and $\sigma_b$ with different bits.
\end{definition}

For example, the elements $(0,1,1)$ and $(1,1,1)$ have Hamming
distance $1$.

\subsection{Reversible and quantum circuits}

A logic circuit consists of interconnected logic gates. A classical
logic gate is a function $f:~\{0,1\}^n~\rightarrow~\{0,1\}^m$ with $n$
input bits and $m$ output bits. We define combinational circuit or
irreversible circuit as an acyclic logic circuit, which means that
each instance of the logic gate is used only once.

When a function $f$ is bijective, it has an inverse
function. Therefore, there is a circuit where, for each output value
$y$ of $f$, it produces the value $x$ such that $f(x)=y$.  In this
case we say that the circuit is reversible. A reversible $n$-gate
realizes a bijective function over $\{ 0,1,\ldots,2^n-1\}$.  For any
reversible gate $g$, the gate $g^{-1}$ implements the inverse
transformation.

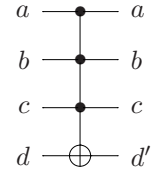
\begin{figure}[ht]
  \centering
\[
\Qcircuit @C=1em @R=1em @!R {
& \lstick{a} & \ctrl{1} & \rstick{a} \qw \\
& \lstick{b} & \ctrl{1} & \rstick{b} \qw \\
& \lstick{c} & \ctrl{1} & \rstick{c} \qw \\
& \lstick{d} & \targ    & \rstick{d'} \qw
}
\]  
\caption{G-Toffoli gate representing $C^4NOT(a,b,c,d)$. The top line
    denotes the less significative bit.}
\label{fig:toffoli:generalized}
\end{figure}

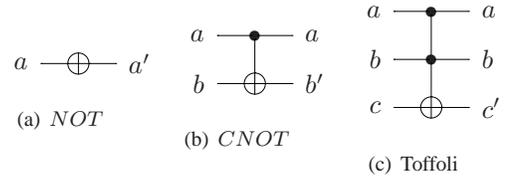
\begin{figure}[ht]
\centering 
\subtable[$NOT$]{\label{fig:toffoli:not}
\parbox{3.0in}{\Qcircuit @C=1em @R=1em @!R {
& \lstick{a} & \targ    & \rstick{a'} \qw \\
&
}}}
\hspace{0.25in}
\subtable[$CNOT$]{\label{fig:toffoli:cnot}
\parbox{3.0in}{
\Qcircuit @C=1em @R=1em @!R {
& \lstick{a} & \ctrl{1} & \rstick{a} \qw \\
& \lstick{b} & \targ    & \rstick{b'} \qw \\
&
}}}
\hspace{0.25in} 
\subtable[Toffoli]{\label{fig:toffoli:toffoli}
\parbox{3.0in}{
\Qcircuit @C=1em @R=1em @!R {
& \lstick{a} & \ctrl{1} & \rstick{a} \qw \\
& \lstick{b} & \ctrl{1} & \rstick{b} \qw \\
& \lstick{c} & \targ    & \rstick{c'} \qw \\
&
}}}
\caption{Circuit representation for the $CNT$ gate library. The top
    line denotes the less significative bit.}
\label{fig:toffoli}
\end{figure}

A \emph{generalized Toffoli gate} or G-Toffoli gate $C^nNOT(x_1,$
$x_2,$ $\ldots,$ $x_n)$ keeps the first $n-1$ lines, called control
lines, unchanged. This gate flips the $n$-th line, target line, if and
only if each control line carries the 1 value. For example,
Figure~\ref{fig:toffoli:generalized} shows a $C^4NOT(a,b,c,d)$
gate. For $n = 0, 1, 2$ the gates are named $NOT (N)$, $CNOT (C)$, and
Toffoli$(T)$, respectively (see Figure~\ref{fig:toffoli}). These three
gates compose the $CNT$ library~\cite{toffoli}, which is a universal
set of gates for the classical reversible computing.

Observe that a reversible $n$-gate applied in a specific position
realizes a permutation of $S_{2^n}$. For example, using decimal
notation, the $NOT$ gate over one line realizes the permutation $[1 \
0]$. If the $NOT$ gate is applied over the most significative bit in a
$2$-line circuit, then the associated permutation is $[2 \ 3 \ 0 \
1]$. The $CNOT$ gate over the $2$-line circuit realizes the
permutation $[0 \ 1 \ 3 \ 2]$ or the permutation $[0 \ 3 \ 2 \ 1]$,
depending on the position of the control bit.

The concatenation of gates in a circuit is equivalent to realizing the
composition of permutations associated to each gate of the
concatenation in the same order.

\begin{definition}
  Let $L$ be a reversible gate library. An \linebreak $L$-circuit is a
  circuit composed only of gates from $L$. A permutation $\pi \in
  S_{2^n}$ is \emph{$L$-constructible} if it can be realized by an
  $L$-circuit.
\end{definition}

\begin{theorem}[Shende et al.~\cite{she}] Every permutation is
  $CNT$-constructible with at most one line of temporary store.
\end{theorem}

\begin{definition}
  \emph{$L_I$} is the reversible gate library formed only by
  generalized Toffoli gates.
\end{definition}

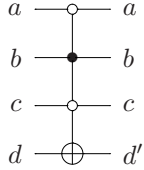
\begin{figure}[ht]
\centering
\[
\Qcircuit @C=1em @R=1em @!R {
& \lstick{a} & \ctrlo{1} & \rstick{a} \qw \\
& \lstick{b} & \ctrl{1} & \rstick{b} \qw \\
& \lstick{c} & \ctrlo{1} & \rstick{c} \qw \\
& \lstick{d} & \targ    & \rstick{d'} \qw
}
\]  
\caption{MC-Toffoli gate representing $C^4NOT(a',b,c',d)$. The top
    line denotes the less significative bit.}
\label{fig:toffoli:mutliple}
\end{figure}

A \emph{multiple-control Toffoli gate} or MC-Toffoli gate
$C^nNOT(x_1,$ $x_2,$ $\ldots,x_{n})$ keeps the first $n-1$ lines,
called control lines, unchanged. This gate flips the $n$-th line,
target line, if and only if each positive (or negative) control line
carries the $1$ (or $0$) value. We indicate the line which is the
negative control with \emph{$'$} after control. See
Figure~\ref{fig:toffoli:mutliple} for an example of a $4 \times 4$
multiple-control Toffoli gate with a negative-positive-negative
pattern of control lines and target on the last line, which can be
denoted by $C^4(a',b,c',d)$.

\begin{theorem}[Toffoli~\cite{toffoli}] Any invertible finite function
  of order $n$ is obtained by the composition of multiple-control
  Toffoli gates.
\end{theorem}

\begin{definition}
  \emph{$L_H$} is the reversible gate library formed only by
  multiple-control Toffoli gates.
\end{definition}

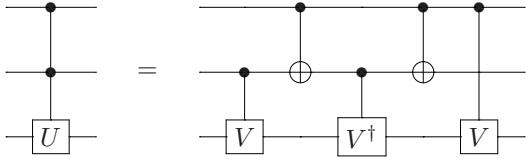
\begin{figure}[ht]
\centering
\[
\Qcircuit @C=1em @R=1em @!R {
& \ctrl{1} & \qw & & & \qw & \ctrl{1} & \qw & \ctrl{1} & \ctrl{2} & \qw\\
& \ctrl{1} & \qw & \push{\rule{.3em}{0em}=\rule{.3em}{0em}} & & \ctrl{1} & \targ & \ctrl{1} & \targ & \qw & \qw\\
& \gate{U} & \qw & & & \gate{V} & \qw & \gate{V^\dag} & \qw & \gate{V} & \qw
}
\]
\caption{Decomposition of two-control quantum gate into a sequence
    of single-control quantum gates. An analog decomposition pattern
    is possible for quantum gates with more than two
    controls~\cite{nielsen00}.}
\label{fig:quant}
\end{figure}

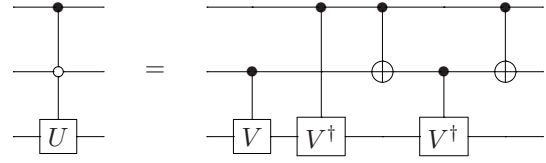
\begin{figure}[ht]
\centering
\[
\Qcircuit @C=1em @R=1em @!R {
& \ctrl{1} & \qw & & & \qw & \ctrl{2} & \ctrl{1} & \qw & \ctrl{1} & \qw\\
& \ctrlo{1} & \qw & \push{\rule{.3em}{0em}=\rule{.3em}{0em}} & & \ctrl{1} & \qw & \targ & \ctrl{1} & \targ & \qw\\
& \gate{U} & \qw & & & \gate{V} & \gate{V^\dag} & \qw & \gate{V^\dag} & \qw & \qw
}
\]  
\caption{Decomposition of two-control quantum gate with one negative
    control into a sequence of single-control quantum gates. An analog
    decomposition pattern is possible for quantum gates with more than
    two controls.}
\label{fig:quant1}
\end{figure}

In the quantum circuit synthesis, a small set of primitive gates are
used as elementary building blocks with an assumed unit
cost~\cite{Barenco1995,mar2,wille1}.  A standard set of universal
gates is composed by Hadamard, phase, CNOT and $\pi/8$
gates~\cite{nielsen00}. In the context of our work, it is also
reasonable to include in this set the NOT gate, the controlled-$V$
gate, and the controlled-$V^\dagger$ gate, with $V$ defined as the
square root of NOT, i.e., a unitary operator such that $V^2$ is equal
to the NOT operator. Each Toffoli gate, $G$-Toffoli gate, or
$MC$-Toffoli gate can be decomposed into a sequence of quantum gates
from the above mentioned set, following the pattern of
Figures~\ref{fig:quant} and~\ref{fig:quant1}.

The number of gates has been used to evaluate nearly all synthesis
approaches in literature so far. For an arbitrary circuit $C$
consisting of a sequence $g_1,g_2,\ldots,g_k$ of $k$ quantum gates,
the \emph{gate count} metric is defined as $gc(C) \equiv k$. We also
refer to the notion of \emph{quantum cost} to measure the
implementation cost of quantum circuits. More precisely, quantum cost
is defined as the number of elementary quantum operations needed to
realize a gate. For an arbitrary quantum gate $g$ that can be
decomposed into $k$ elementary quantum gates, its \emph{quantum cost}
metric is defined as $qc(g) \equiv k$. The quantum cost for a circuit
$C$ is defined as $qc(C) = \sum_{g_i \in C}
qc(g_i)$. Table~\ref{tab:cost} shows the quantum cost for all the
reversible gates used in this paper, with $m$ denoting the amount of
negative controls on the $MC$-Toffoli gate.

\begin{table}
  \begin{center}
    \caption{Quantum costs of various gates}
    \label{tab:cost}
    \begin{tabular}{|c|c|c|c|}
      \hline
      gate type & size & garbage & quantum cost  \\		
      \hline
      $NOT$ & 1 & 0 & 1~\cite{nielsen00} \\
      $CNOT$ & 2 & 0 & 1~\cite{nielsen00}\\
      Toffoli & 3 & 0 & 5~\cite{Barenco1995}\\
      Toffoli gate  & 3 & 0 & 5\\
      with one negative control & & & \\
      Toffoli gate  & 3 & 0 & 7\\        
      with two negative control & & &  \\
      \hline
      $G$-Toffoli & & & \\
      $C^nNOT(x_1,x_2,\ldots,x_n)$ & $n$ & 0 & $2^n - 3$~\cite{Barenco1995}\\   
      $C^nNOT(x_1,x_2,\ldots,x_n)$ & $n$ & 1 & $24n - 88$~\cite{Barenco1995,mar2} \\   
      $C^nNOT(x_1,x_2,\ldots,x_n)$ & $n$ & $n-3$ & $10n - 25$~\cite{luis,dmi} \\
      \hline
      $MC$-Toffoli & & & \\
      $C^nNOT(x_1,x_2,\ldots,x_n)$ & $n$ & 0 & $2^n - 3 + 2m$ \\   
      $C^nNOT(x_1,x_2,\ldots,x_n)$ & $n$ & 1 & $24n - 86 $ \\   
      $C^nNOT(x_1,x_2,\ldots,x_n)$ & $n$ & $n-3$ & $10n - 23$ \\   
      \hline		
    \end{tabular}
  \end{center}
\end{table}

\begin{figure}[ht]
\centering
\[
\Qcircuit @C=0.8em @R=0.6em @!R {
& \lstick{a} & \ctrl{1} & \rstick{a} \qw & & \lstick{a} & \qw & \ctrl{1} & \qw & \qw & \qw & \qw & \qw & \ctrl{1} & \rstick{a} \qw \\
& \lstick{b} & \ctrlo{1} & \rstick{b} \qw & & \lstick{b} & \qw & \ctrlo{4} & \qw & \qw & \qw & \qw & \qw & \ctrlo{4} & \rstick{b} \qw \\
& \lstick{c} & \ctrlo{1} & \rstick{c} \qw & & \lstick{c} & \qw & \qw & \ctrlo{3} & \qw & \qw & \qw & \ctrlo{3} & \qw & \rstick{c} \qw \\
& \lstick{d} & \ctrlo{1} & \rstick{d} \qw & & \lstick{d} & \qw & \qw & \qw & \ctrlo{3} & \qw & \ctrlo{3} & \qw & \qw & \rstick{d} \qw \\
& \lstick{e} & \ctrlo{4} & \rstick{e} \qw & \push{\rule{.3em}{0em}=\rule{.3em}{0em}} & \lstick{e} & \qw & \qw & \qw & \qw & \ctrlo{3} & \qw & \qw & \qw & \rstick{e} \qw \\
& \lstick{0} & \qw & \rstick{0} \qw &  & \lstick{0} & \qw & \targ & \ctrl{1} & \qw & \qw & \qw & \ctrl{1} & \targ & \rstick{0} \qw \\
& \lstick{0} & \qw & \rstick{0} \qw &  & \lstick{0} & \qw & \qw & \targ & \ctrl{1} & \qw & \ctrl{1} & \targ & \qw & \rstick{0} \qw \\
& \lstick{0} & \qw & \rstick{0} \qw &  & \lstick{0} & \qw & \qw & \qw & \targ & \ctrl{1} & \targ & \qw & \qw & \rstick{0} \qw \\
& \lstick{f} & \targ & \rstick{f'} \qw & & \lstick{f} & \qw & \qw & \qw & \qw & \targ & \qw & \qw & \qw & \rstick{f'} \qw
}
\]
\caption{Implementation of MC-Toffoli for $n=6$ and $3$ garbage based on~\cite{luis,dmi}.}
\label{fig:n6}
\end{figure}
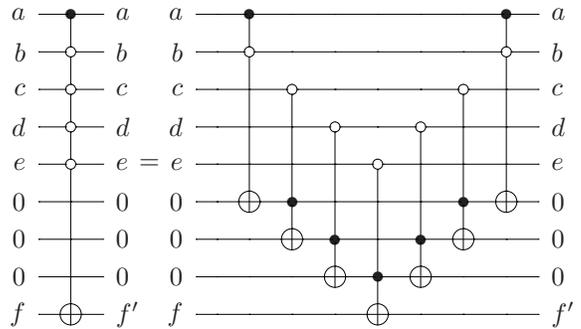

In Figure~\ref{fig:n6}, we have an example of how to decompose a
MC-Toffoli gate of size $n=6$ into Toffoli gates, with $n-3$ ancilla
(garbage) bits, by using a synthesis method based on~\cite{luis,dmi}. Quantum cost, in this case, is the gate count
multiplied by $5$.

\begin{figure}[ht]
\centering
\[
\Qcircuit @C=0.5em @R=0.6em @!R {
& \lstick{a} & \ctrl{1} & \rstick{a} \qw & & \lstick{a} & \qw & \qw & \qw & \qw & \ctrl{1} & \qw & \qw & \qw & \qw & \qw & \ctrl{1} & \qw & \qw & \rstick{a} \qw \\
& \lstick{b} & \ctrlo{1} & \rstick{b} \qw & & \lstick{b} & \qw & \qw & \qw & \qw & \ctrlo{4} & \qw & \qw & \qw & \qw & \qw & \ctrlo{4} & \qw & \qw & \rstick{b} \qw \\
& \lstick{c} & \ctrlo{1} & \rstick{c} \qw & & \lstick{c} & \qw & \qw & \qw & \ctrlo{3} & \qw & \ctrlo{3} & \qw & \qw & \qw & \ctrlo{3} & \qw & \ctrlo{3} & \qw & \rstick{c} \qw \\
& \lstick{d} & \ctrlo{1} & \rstick{d} \qw & & \lstick{d} & \qw & \qw & \ctrlo{3} & \qw & \qw & \qw & \ctrlo{3} & \qw & \ctrlo{3} & \qw & \qw &\qw & \ctrlo{3} & \rstick{d} \qw \\
& \lstick{e} & \ctrlo{4} & \rstick{e} \qw & \push{\rule{.3em}{0em}=\rule{.3em}{0em}} & \lstick{e} & \qw & \ctrlo{3} & \qw & \qw & \qw & \qw & \qw & \ctrlo{3} & \qw & \qw & \qw & \qw & \qw & \rstick{e} \qw \\
& \lstick{x} & \qw & \rstick{x} \qw &  & \lstick{x} & \qw & \qw & \qw & \ctrl{1} & \targ & \ctrl{1} & \qw & \qw & \qw & \ctrl{1} & \targ & \ctrl{1} & \qw & \rstick{x} \qw \\
& \lstick{y} & \qw & \rstick{y} \qw &  & \lstick{y} & \qw & \qw & \ctrl{1} & \targ & \qw & \targ & \ctrl{1} & \qw & \ctrl{1} & \targ & \qw & \targ & \ctrl{1} & \rstick{y} \qw \\
& \lstick{z} & \qw & \rstick{z} \qw &  & \lstick{z} & \qw & \ctrl{1} & \targ & \qw & \qw & \qw & \targ & \ctrl{1} & \targ & \qw & \qw & \qw & \targ & \rstick{z} \qw \\
& \lstick{f} & \targ & \rstick{f'} \qw & & \lstick{f} & \qw & \targ & \qw & \qw & \qw & \qw & \qw & \targ & \qw & \qw & \qw & \qw & \qw & \rstick{f'} \qw
}
\]
\caption{Implementation of MC-Toffoli for $n=6$ and $3$ garbage based on~\cite{Barenco1995,mar2}.}
\label{fig:n6a}
\end{figure}
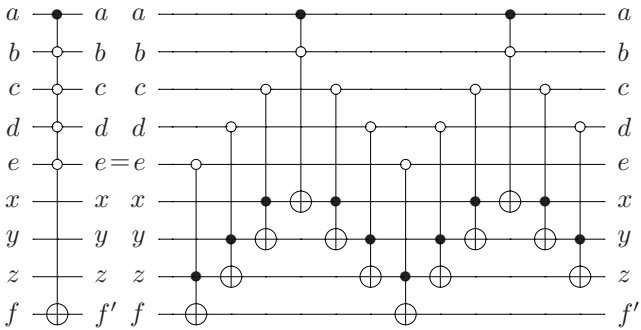

In Figure~\ref{fig:n6a}, we have an example of how to decompose a
MC-Toffoli gate of size $n=6$ into Toffoli gates, with $n-3$ garbage
bits, by using a synthesis method based on~\cite{Barenco1995,mar2}. Notice that in this case, we do not need
to initialize the garbage bits with zeros.

In Figures~\ref{fig:n8} and~\ref{fig:n8a}, we have an example of how
to decompose a MC-Toffoli gate of size $n=8$ into Toffoli gates, with
one garbage bit, by using a synthesis method based on~\cite{Barenco1995,mar2}.

\begin{figure}[ht]
\centering
\[
\Qcircuit @C=0.8em @R=0.6em @!R {
& \lstick{a} & \ctrl{1} & \rstick{a} \qw & & \lstick{a} & \qw & \ctrl{1} & \qw & \ctrl{1} & \qw & \rstick{a} \qw \\
& \lstick{b} & \ctrlo{1} & \rstick{b} \qw & & \lstick{b} & \qw & \ctrlo{1} & \qw & \ctrlo{1} & \qw & \rstick{b} \qw \\
& \lstick{c} & \ctrlo{1} & \rstick{c} \qw & & \lstick{c} & \qw & \ctrlo{1} & \qw & \ctrlo{1} & \qw &  \rstick{c} \qw \\
& \lstick{d} & \ctrlo{1} & \rstick{d} \qw & & \lstick{d} & \qw & \ctrlo{1} & \qw & \ctrlo{1} & \qw &  \rstick{d} \qw \\
& \lstick{e} & \ctrlo{1} & \rstick{e} \qw & \push{\rule{.3em}{0em}=\rule{.3em}{0em}} & \lstick{e} & \qw & \ctrlo{3} & \qw & \ctrlo{3} & \qw & \rstick{e} \qw \\
& \lstick{f} & \ctrlo{1} & \rstick{f} \qw &  & \lstick{f} & \qw & \qw & \ctrlo{1} & \qw & \ctrlo{1} &  \rstick{f} \qw \\
& \lstick{g} & \ctrlo{2} & \rstick{g} \qw &  & \lstick{g} & \qw & \qw & \ctrlo{1} & \qw & \ctrlo{1} &  \rstick{g} \qw \\
& \lstick{h} & \qw & \rstick{h} \qw &  & \lstick{h} & \qw & \targ & \ctrl{1} & \targ & \ctrl{1} &  \rstick{h} \qw \\
& \lstick{i} & \targ & \rstick{i'} \qw & & \lstick{i} & \qw & \qw & \targ & \qw & \targ & \rstick{i'} \qw
}
\]
\caption{Implementation of MC-Toffoli for $n=8$ and $1$ garbage based on~\cite{Barenco1995,mar2}.}
\label{fig:n8}
\end{figure}

\begin{figure}[ht]
\centering
\[
\Qcircuit @C=-0.1em @R=0.2em @!R {
& \lstick{a} & \qw & \qw & \qw & \qw & \ctrl{1} & \qw & \qw & \qw & \qw & \qw & \ctrl{1} & \qw & \qw & \qw & \qw & \qw & \qw & \qw & \qw & \qw & \ctrl{1} & \qw & \qw & \qw & \qw & \qw & \ctrl{1} & \qw & \qw & \qw & \qw & \qw & \qw & \rstick{a} \qw \\
& \lstick{b} & \qw & \qw & \qw & \qw & \ctrlo{4} & \qw & \qw & \qw & \qw & \qw & \ctrlo{4} & \qw & \qw & \qw & \qw & \qw & \qw & \qw & \qw & \qw & \ctrlo{4} & \qw & \qw & \qw & \qw & \qw & \ctrlo{4} & \qw & \qw & \qw & \qw & \qw & \qw & \rstick{b} \qw \\
& \lstick{c} & \qw & \qw & \qw & \ctrlo{3} & \qw & \ctrlo{3} & \qw & \qw & \qw & \ctrlo{3} & \qw & \ctrlo{3} & \qw & \qw & \qw & \qw & \qw & \qw & \qw & \ctrlo{3} & \qw & \ctrlo{3} & \qw & \qw & \qw & \ctrlo{3} & \qw & \ctrlo{3} & \qw & \qw & \qw & \qw & \qw & \rstick{c} \qw \\
& \lstick{d} & \qw & \qw & \ctrlo{3} & \qw & \qw & \qw & \ctrlo{3} & \qw & \ctrlo{3} & \qw & \qw &\qw & \ctrlo{3} & \qw & \qw & \qw & \qw & \qw & \ctrlo{3} & \qw & \qw & \qw & \ctrlo{3} & \qw & \ctrlo{3} & \qw & \qw &\qw & \ctrlo{3} & \qw & \qw & \qw & \qw & \rstick{d} \qw \\
& \lstick{e} & \qw & \ctrlo{3} & \qw & \qw & \qw & \qw & \qw & \ctrlo{3} & \qw & \qw & \qw & \qw & \qw & \targ & \ctrl{1} & \targ & \ctrl{1} & \ctrlo{3} & \qw & \qw & \qw & \qw & \qw & \ctrlo{3} & \qw & \qw & \qw & \qw & \qw & \targ & \ctrl{1} & \targ & \ctrl{1} & \rstick{e} \qw \\
& \lstick{f} & \qw & \qw & \qw & \ctrl{1} & \targ & \ctrl{1} & \qw & \qw & \qw & \ctrl{1} & \targ & \ctrl{1} & \qw & \qw & \ctrlo{3} & \qw & \ctrlo{3} & \qw & \qw & \ctrl{1} & \targ & \ctrl{1} & \qw & \qw & \qw & \ctrl{1} & \targ & \ctrl{1} & \qw & \qw & \ctrlo{3} & \qw & \ctrlo{3} & \rstick{f} \qw \\
& \lstick{g} & \qw & \qw & \ctrl{2} & \targ & \qw & \targ & \ctrl{2} & \qw & \ctrl{2} & \targ & \qw & \targ & \ctrl{2} & \ctrlo{-2} & \qw & \ctrlo{-2}  & \qw & \qw & \ctrl{2} & \targ & \qw & \targ & \ctrl{2} & \qw & \ctrl{2} & \targ & \qw & \targ & \ctrl{2} & \ctrlo{-2} & \qw & \ctrlo{-2}  & \qw & \rstick{g} \qw \\
& \lstick{h} & \qw & \targ & \qw & \qw & \qw & \qw & \qw & \targ & \qw & \qw & \qw & \qw & \qw & \ctrl{-1}  & \qw & \ctrl{-1}  & \qw & \targ & \qw & \qw & \qw & \qw & \qw & \targ & \qw & \qw & \qw & \qw & \qw & \ctrl{-1}  & \qw & \ctrl{-1}  & \qw & \rstick{h} \qw \\
& \lstick{i} & \qw & \ctrl{-1} & \targ & \qw & \qw & \qw & \targ & \ctrl{-1} & \targ & \qw & \qw & \qw & \targ & \qw & \targ & \qw & \targ & \ctrl{-1} & \targ & \qw & \qw & \qw & \targ & \ctrl{-1} & \targ & \qw & \qw & \qw & \targ & \qw & \targ & \qw & \targ & \rstick{i'} \qw
}
\]
\caption{Implementation of MC-Toffoli for $n=8$ and $1$ garbage based on~\cite{Barenco1995,mar2}.}
\label{fig:n8a}
\end{figure}

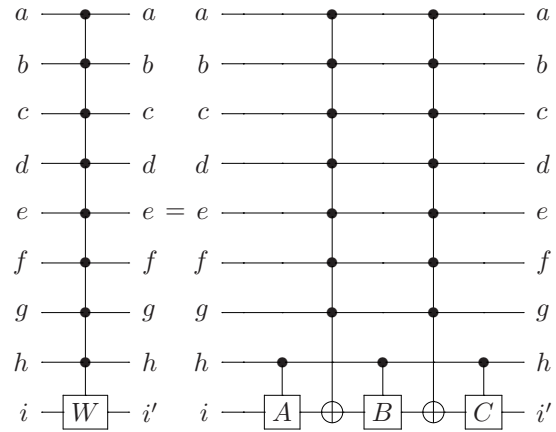
\begin{figure}[ht]
\centering
\[
\Qcircuit @C=0.8em @R=0.6em @!R {
& \lstick{a} & \ctrl{1} & \rstick{a} \qw & & \lstick{a} & \qw & \qw & \ctrl{1} & \qw & \ctrl{1} & \qw & \rstick{a} \qw \\
& \lstick{b} & \ctrl{1} & \rstick{b} \qw & & \lstick{b} & \qw & \qw & \ctrl{1} & \qw & \ctrl{1} & \qw & \rstick{b} \qw \\
& \lstick{c} & \ctrl{1} & \rstick{c} \qw & & \lstick{c} & \qw & \qw & \ctrl{1} & \qw & \ctrl{1} & \qw &  \rstick{c} \qw \\
& \lstick{d} & \ctrl{1} & \rstick{d} \qw & & \lstick{d} & \qw & \qw & \ctrl{1} & \qw & \ctrl{1} & \qw &  \rstick{d} \qw \\
& \lstick{e} & \ctrl{1} & \rstick{e} \qw & \push{\rule{.3em}{0em}=\rule{.3em}{0em}} & \lstick{e} & \qw & \qw & \ctrl{1} & \qw & \ctrl{1} & \qw & \rstick{e} \qw \\
& \lstick{f} & \ctrl{1} & \rstick{f} \qw &  & \lstick{f} & \qw & \qw & \ctrl{1} & \qw & \ctrl{1} & \qw &  \rstick{f} \qw \\
& \lstick{g} & \ctrl{1} & \rstick{g} \qw &  & \lstick{g} & \qw & \qw  & \ctrl{2} & \qw & \ctrl{2} & \qw & \rstick{g} \qw \\
& \lstick{h} & \ctrl{1} & \rstick{h} \qw &  & \lstick{h} & \qw & \ctrl{1} & \qw & \ctrl{1} & \qw & \ctrl{1} &  \rstick{h} \qw \\
& \lstick{i} & \gate{W} & \rstick{i'} \qw & & \lstick{i} & \qw & \gate{A} & \targ & \gate{B} & \targ & \gate{C} & \rstick{i'} \qw 
}
\]
\caption{G-Toffoli n=9 and 0 garbage by~\cite{Barenco1995,mar2}.}
\label{fig:u9}
\end{figure}

In Figure~\ref{fig:u9}, we have an example of how to decompose a
G-Toffoli gate of size $n=9$ into Toffoli gates, without garbage bits,
by using a synthesis method based on~\cite{Barenco1995,mar2}.

\subsection{Circuit synthesis using generalized Toffoli gates}
\label{subgeneralized}

We show below the reversible circuit synthesis using generalized
Toffoli gate, which we call \emph{Basic Algorithm}. The Basic
Algorithm was proposed by Maslov, Dueck and Miller~\cite{mar,mar1},
and is reproduced in Algorithm~\ref{alg:basic}. It considers a
reversible function specified as a mapping over $\{ 0,
1,\ldots,{2^n-1}\}$, i.e., a truth vector. It writes a
function $f(i)$, where $i$ is an integer in the range $0 \leq i \leq
2^n-1$, meaning that the function argument $i$ is a vector giving the
binary expansion of the integer $i$. The result of the function
application to an integer argument $i$, $f(i)$, is treated as an
integer as well. The Basic Algorithm works by assigning Toffoli gates
at the output end of the cascade. The Toffoli gates are chosen so that
the output part of the specification is progressively transformed to
match the input part. When a cascade of Toffoli gates transforming the total specification into the identity permutation is found, then reading this cascade in reverse order will transform the input to the required output, thus realizing the target function.

\begin{center}
\begin{algorithm}
\Begin {

\textbf{Step 0:} \textit{If $f(0) \notin 0$, invert the outputs corresponding to $1$-bits in $f(0)$. Each inversion requires a NOT gate. The transformed function, written as $f^+$, has $f^+(0)=0$.}

\textbf{Step i:} \textit{Consider each $i$ in turn for $0 \leq i \leq 2^n-1$ letting $f^+$ denote the current reversible specification. If $f^+(i)=i$, no transformation and, hence, no Toffoli gate is required for this $i$. Otherwise, gates are required to transform the specification to a new specification $f^{++}$ with $f^{++}(i)=i$. The required gates must map $f^{+}(i) \rightarrow i$.}

\textit{Let $p$ be the bit string with $1s$ in all position where the binary expansion of $i$ is $1$, while the expansion of $f^+(i)$ is $0$. These are the $1$ bits that must be added in transforming $f^+(i) \rightarrow i$. Conversely, let $q$ be the bit string with $1s$ in all positions where the expansion of $i$ is $0$, while the expansion of $f^+(i)$ is $1$. $q$ identifies the $1$ bits to be removed in the transformation.}

\textit{For each $p_j=1$, apply the Toffoli gate with control lines corresponding to all outputs in positions where the expansion of $i$ is $1$ and whose target line is the output in position $j$. This will increase the lexicographical order of $f^+(i)$. Then, for each $q_k=1$, apply the Toffoli gate whose target line is the output in position $k$, and with control lines corresponding to all outputs in positions, except $k$, where the expansion of $f^+(i)$ is $1$. This second operation decreases the lexicographical order, but not below $i$.}

}
\caption{Basic Algorithm~\cite{mar,mar1}}
\label{alg:basic}
\end{algorithm}
\end{center}

\begin{table}
  \begin{center}
    \caption{Example of applying the Basic Algorithm}
    \label{tab:exbasicalgorithm}
    \begin{tabular}{|c|c|c|c|c|c|}
      \hline
      & (i) & (ii) & (iii) & (iv) & (v) \\		
      \hline
      cba & $c^0b^0a^0$ & $c^1b^1a^1$ & $c^2b^2a^2$ & $c^3b^3a^3$ & $c^4b^4a^4$ \\
      \hline
      000 & 001 & 000 & 000 & 000 & 000 \\
      001 & 000 & 001 & 001 & 001 & 001 \\
      010 & 011 & 010 & 010 & 010 & 010 \\
      011 & 010 & 011 & 011 & 011 & 011 \\
      100 & 101 & 100 & 100 & 100 & 100 \\
      101 & 111 & 110 & 111 & 101 & 101 \\
      110 & 100 & 101 & 101 & 111 & 110 \\
      111 & 110 & 111 & 110 & 110 & 111 \\    
      \hline		
    \end{tabular}
  \end{center}
\end{table}

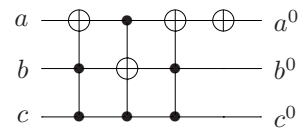
\begin{figure}[htp]
\centering
\[
\Qcircuit @C=1em @R=1em @!R {
& \lstick{a} & \targ     & \ctrl{1}  & \targ     & \targ & \rstick{a^0} \qw \\
& \lstick{b} & \ctrl{-1} & \targ     & \ctrl{-1} & \qw & \rstick{b^0} \qw \\
& \lstick{c} & \ctrl{-1} & \ctrl{-1} & \ctrl{-1} & \qw & \rstick{c^0} \qw \\
}
\]
\caption{\label{fig:basic} Reversible circuit that transforms
    permutation $\iota$ into $\pi$, according to example taken from
    Table~\ref{tab:exbasicalgorithm}.}
\end{figure}

Table~\ref{tab:exbasicalgorithm} illustrates the application of the
Basic Algorithm. Notice that the gates are identified in order from
the output side to the input side. The corresponding network is showed
in Figure~\ref{fig:basic}.

Using the Basic Algorithm, it is possible to find a permutation for
any $n$ that requires at most $(n-1)2^n + 1$ generalized Toffoli
gates. The Basic Algorithm finds the permutation $[7 \ 1 \ 4 \ 3 \ 0 \
2 \ 6 \ 5]$ for $n=3$ and the permutation $[15 \ 1 \ 12 \ 3 \ 5 \ 6 \
8 \ 7 \ 0 \ 10 \ 13 \ 9 \ 2 \ 4 \ 14 \ 11]$ for $n=4$.

\begin{table}[ht]
  \caption{Number of reversible $3 \times 3$ functions for Algorithm~\ref{alg:basic} using $G$-Toffoli gates and comparing with optimal results from~\cite{she}. For each scenario, we show the average number of $G$-Toffoli gates required.}
  \label{tab:n3basic}
  \begin{center}
    \begin{tabular}{|c|c|c|}
      \hline
      \multirow{2}{*}{Size} &  \multicolumn{2}{c|}{Number of permutations} \\
      \cline{2-3}
      & Algorithm~\ref{alg:basic}  & optimal results  \\
      &   &  by $CNT$ library \\      \hline
      17 & 1 & \\
      16 & 14 & \\
      15 & 92	& \\
      14 & 380 & \\
      13 & 1113	& \\
      12 & 2468 & \\
      11 & 4311 & \\
      10 & 6083 & \\
      9 & 7044 & \\
      8 & 6754 & 577\\
      7 & 5379 & 10253 \\
      6 & 3549 & 17049\\
      5 & 1922 & 8921 \\
      4 & 839 & 2780 \\
      3 & 286 & 625 \\
      2 & 72 & 102 \\
      1 & 12 & 12 \\
      0 & 1  & 1 \\
      \hline
      avg. gates & 8.67 & 5.63 \\
      \hline
    \end{tabular}
  \end{center}
\end{table}

Table~\ref{tab:n3basic} shows the results of applying version of
Algorithm~\ref{alg:basic} over all $8!=40320$ permutations when $n=3$. We show the number of functions for each gate count and the average number of gates required. 

\subsection{Circuit synthesis using multiple-control Toffoli gates}
\label{submultiple}

We present below the reversible circuit synthesis using
multiple-control Toffoli gates, denoted by \emph{hypercube method}. Each permutation is a sequence of $n2^n$ bits. The application of a multiple-control Toffoli gate over one permutation $\pi$
generates a permutation $\pi'$, with change over two bits, i.e.,
$d_H(\pi,\pi')=2$. Regarding the identity permutation $\iota$, we have
three cases for the Hamming distance: i)
$d_H(\pi_b,\iota_b)=d_H(\pi'_b,\iota_b)-2$, when the gate places two
bits in their correct positions; ii)
$d_H(\pi_b,\iota_b)=d_H(\pi'_b,\iota_b)$, when the gate places one bit
in its correct position and misplaces one bit in a wrong position;
iii) $d_H(\pi_b,\iota_b)=d_H(\pi'_b,\iota_b)+2$, when the gate
misplaces two bits in wrong positions.

\begin{table}
  \scriptsize
  \begin{center}
    \caption{The evolution of permutation $\pi$ being transformed into
      $\iota$ by hypercube method. The elements that need to be moved
      are presented in bold typeface. Ordered elements are presented
      underlined.}
    \label{tab:conjuntosrp}
    \begin{tabular}{|p{19mm}|c|c|c|c|c|c|c|c|}
      \hline
      & \multicolumn{8}{|c|}{permutation elements}\\
      \hline
      apply gate & 7 & 4 & 1 & 0 & 3 & 2 & 6 & 5 \\
      \hline
      & 111 & 100 & 001 & 000 & 011 & 010 & 110 & 101 \\		
      \hline
      step $i=7$ & \textbf{111} & 100 & 001 & 000 & 011 & 010 & 110 & \textbf{101} \\
      $C^3NOT(a,c,b)$ &&&&&&&& \\
      & 101 & 100 & 001 & 000 & 011 & 010 & 110 & \underline{111} \\				
      \hline
      step $i=6$ & 101 & 100 & 001 & 000 & 011 & 010 & \underline{110} & \underline{111} \\				
      \hline
      step $i=5$ & 101 & 100 & 001 & 000 & \textbf{011} & \textbf{010} & \underline{110} & \underline{111} \\				
      $C^3NOT(b,c',a)$ &&&&&&&& \\		
      & 101 & 100 & \textbf{001} & 000 & 010 & \textbf{011} & \underline{110} & \underline{111} \\		
      $C^3NOT(a,c',b)$ &&&&&&&& \\				
      & \textbf{101} & 100 & 011 & 000 & 010 & \textbf{001} & \underline{110} & \underline{111} \\
      $C^3NOT(a,b',c)$ &&&&&&&& \\										
      &001 & 100 & 011 & 000 & 010 & \underline{101} & \underline{110} & \underline{111} \\  	
      \hline	
      step $i=4$ &001 & 100 & 011 & \textbf{000} & \textbf{010} & \underline{101} & \underline{110} & \underline{111} \\
      $C^3NOT(a',c',b)$ &&&&&&&& \\
      &001 & \textbf{100} & 011 & 010 & \textbf{000} & \underline{101} & \underline{110} & \underline{111} \\
      $C^3NOT(a',b',c)$ &&&&&&&& \\  													  	
      &001 & 000 & 011 & 010 & \underline{100} & \underline{101} & \underline{110} & \underline{111} \\  
      \hline
      step $i=3$ &001 & 000 & \textbf{011} & \textbf{010} & \underline{100} & \underline{101} & \underline{110} & \underline{111} \\  	
      $C^3NOT(b,c',a)$ &&&&&&&& \\  													  	
      &001 & 000 & 010 & \underline{011} & \underline{100} & \underline{101} & \underline{110} & \underline{111} \\   	
      \hline
      step $i=2$ &001 & 000 & \underline{010} & \underline{011} & \underline{100} & \underline{101} & \underline{110} & \underline{111} \\   	
      \hline												
      step $i=1$ &\textbf{001} & \textbf{000} & \underline{010} & \underline{011} & \underline{100} & \underline{101} & \underline{110} & \underline{111} \\   	
      $C^3NOT(b',c',a)$ &&&&&&&& \\  													  	
      &\underline{000} & \underline{001} & \underline{010} & \underline{011} & \underline{100} & \underline{101} & \underline{110} & \underline{111} \\    
      \hline		
      &\underline{0} & \underline{1} & \underline{2} & \underline{3} & \underline{4} & \underline{5} & \underline{6} & \underline{7} \\    	
      \hline		
    \end{tabular}
  \end{center}
\end{table}

\begin{figure}[htp]
\centering
\[
\Qcircuit @C=1em @R=1em @!R {
& \lstick{a} & \targ      & \targ      & \ctrlo{1} & \ctrlo{1}  & \ctrl{1}   & \ctrl{1}   & \targ & \ctrl{1}  & \rstick{a'} \qw \\
& \lstick{b} & \ctrlo{-1} & \ctrl{-1}  & \ctrlo{1} & \targ      & \ctrlo{1}  & \targ      & \ctrl{-1} & \targ  & \rstick{b'} \qw \\
& \lstick{c} & \ctrlo{-1} & \ctrlo{-1} & \targ     & \ctrlo{-1} & \targ      & \ctrlo{-1} & \ctrlo{-1} & \ctrl{-1}  & \rstick{c'} \qw \\
}
\]
\caption{\label{fig:p74103265r} Reversible circuit that transforms
    permutation $\iota$ into $\pi$, according to example taken from
    Table~\ref{tab:conjuntosrp}.}
\end{figure}
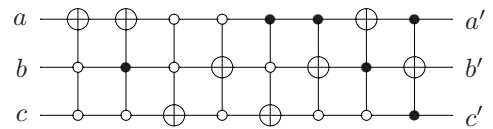

The hypercube method for reversible circuit synthesis uses the $L_H$
gate library. This method, presented in the Algorithm~\ref{alg:hyper}, uses consecutive applications of multiple-control Toffoli gates in order to organize the bits. The proposed method takes the binary representation of the permutation elements---each one is composed of $n$ bits---and carries out at most $n$ changes. Those changes use the multiple-control Toffoli gate in order to put the permutation element in its correct position. See
example in Table~\ref{tab:conjuntosrp}. The corresponding reversible
circuit is given in Figure~\ref{fig:p74103265r}.

For instance, Table~\ref{tab:conjuntosrp} shows the hypercube method
that transform the permutation $\pi=[ 7 \ 4 \ 1 \ 0 \ 3 \ 2 \ 6 \ 5]$
into the identity permutation $\iota = [ 0 \ 1 \ 2 \ 3 \ 4 \ 5 \ 6 \
7]$.

Algorithm~\ref{alg:hyper} makes the reversible circuit synthesis using
multiple-control Toffoli gates and reads the permutation from right to
left order. We call it the \emph{right order}. One can change the line
$4$ of Algorithm~\ref{alg:hyper} to read the permutation from left to
right order. In that case, we call it the \emph{left order}. Denoted by \emph{unidirectional} if Algorithm~\ref{alg:hyper} runs right order or left order. Denoted by \emph{bidirectional} if Algorithm~\ref{alg:hyper} runs right order and left order.

\begin{center}
  \begin{algorithm}
    \SetKwInOut{Input}{input} \SetKwInOut{Output}{output}
    \Input{$\pi_b$ vector} \Output{$circuit$ stack}

\Begin {

$N = $ lenght of $\pi_b$ vector\;
$n = \log_2 N$\;

\For{$i=N-1$ until $1$}
{
\If{$\pi_b[i] \neq \iota_b[i]$}
{
\For{$j=n-1$ until $0$}
{
\If{$\pi_b[i][j] \neq \iota_b[i][j]$}
{
\tcc{Adds to the circuit a MC-Toffoli with target in position $j$ and controls in positions $\pi_b - j$}
push $C^nNOT(\pi_b[0],\ldots,\pi_b[n-1],\pi_b[j])$ to $circuit$\; 
change $\pi_b[i][j]$\;
}}}}

\Return{$circuit$}

}
\caption{Reversible circuit synthesis}
\label{alg:hyper}
\end{algorithm}
\end{center}

The hypercube method has the following property in the change of
permutation and it is used to prove the correctness of Algorithm~\ref{alg:hyper}.

\begin{property}
  When $i < k$ in the external for loop in Algorithm~\ref{alg:hyper},
  let $\pi$ be any permutation and let $\pi[k]$ be a component of
  permutation $\pi$, where $i < k \leq 2^{n}-1$. Then we have that
  $\pi[k] = k$.
  \label{pro:pro1sr}
\end{property}

\begin{theorem}
  Algorithm~\ref{alg:hyper} returns a reversible circuit using
  multiple-control Toffoli gates.
\end{theorem}

\begin{IEEEproof}
  We will prove by induction the correctness of the
  Algorithm~\ref{alg:hyper} by showing that in the application of each
  multiple-control Toffoli gate, at least one bit changes or stays in
  its correct place.  The induction hypothesis: In the step $i=k$ and
  $j=l$, we have $\pi[k']=k'$ and $\pi_b[k][l']=\iota_b[k][l']$ for
  $k'> k$ and $l'>l$, or in other words, the elements greater than $k$
  and all less significative bits than $\pi_b[k][l]$ of $\pi[k]$ stay
  in their correct positions.

  When $i=N-1$, it is only possible to change $\pi[i]$ with $\pi[k']$,
  where $k'<N-1$, warranting the satisfiability of the basis of
  induction.

  The induction is guaranteed if we show that the gate application, in
  the step $i=k$, does not affect any element $\pi[k']$, with $k'>k$.

  In the step $i=k$ and $j=l$, if $\pi_b[k][l]=\iota_b[k][l]$ then no
  change is needed. Otherwise, $\pi[k]$ must change the $l$-th bit
  with $\pi[m]$ such that
$$
\pi_b[m][p]= \left\{
  \begin{array}{ll}
    \pi_b[k][p] & \textrm{if } p\ne l\\
    !\pi_b[k][p] & \textrm{if } p = l,\\
  \end{array}
\right.
$$
where $!x$ is $1$ if $x=0$ or $0$ in otherwise.

If $\pi_b[k][l]=1$ then the change $(1$ to $0)$ is made to a value
$\pi[m]$ less than $\pi[k]$, consequently, by induction hypothesis, to
a value less than $k$.  If $\pi_b[k][l]=0$ then the change is made to
a value greater than $\pi[k]$. We must show that the $\pi[m]<k$.

Take the most significative bit $l'$ of $\pi[k]$ such that
$\pi_b[k][l']\ne\iota_b[k][l']$. We can separate in two cases: (i)
$l'=l$ (ii) $l'>l$.  In the case (i), the bits more significative than
$l$ of $\pi[k]$ are correct and by induction hypothesis, the bits less
significative than $l$ of $\pi[k]$ are correct and $\pi[k']=k'$ for
$k'> k$, hence if $m>k$ then $\pi[m]=k$ and $\pi[m]=k'$, that is not
possible.  In the case (ii), if $m>k$ then $\pi_b[m][l']=1$, but
$\pi_b[m][l']=\pi_b[k][l']$ when $l'\ne l$, thus $\pi[k]>k$ that is
contrary to induction hypothesis.
\end{IEEEproof}

\begin{table}[ht]
  \caption{Number of reversible $3 \times 3$ functions for Algorithm~\ref{alg:hyper} using $MC$-Toffoli gates and  comparing with optimal results from~\cite{she}. For each scenario, we show the average number of $MC$-Toffoli gates required.}
  \label{tab:n3}
  \begin{center}
    \begin{tabular}{|c|c|c|c|}
      \hline
      \multirow{2}{*}{Size} &  \multicolumn{3}{c|}{Number of permutations} \\
      \cline{2-4}
      & Algorithm~\ref{alg:hyper} & Algorithm~\ref{alg:hyper}  & optimal results  \\
      &   unidirectional  & bidirectional &  by $NT$ library \\
      \hline
      17 & 1 & & \\
      16 & 14 & & \\
      15 & 92	& & \\
      14 & 380 & 9 & \\
      13 & 1113	& 111 & \\
      12 & 2468 & 581 & 47 \\
      11 & 4311 & 1946 & 1690 \\
      10 & 6083 & 4349 & 8363 \\
      9 & 7044 & 6917 & 12237 \\
      8 & 6754 & 8255 & 9339 \\
      7 & 5379 & 7662 & 5097 \\
      6 & 3549 & 5546 & 2262 \\
      5 & 1922 & 3088 & 870 \\
      4 & 839 & 1329 & 296 \\
      3 & 286 & 424 & 88 \\
      2 & 72 & 90 & 24 \\
      1 & 12 & 12 & 6 \\
      0 & 1  & 1 & 1 \\
      \hline
      avg. gates & 8.67 & 7.71 & 8.50 \\
      \hline
    \end{tabular}
  \end{center}
\end{table}

\begin{theorem}
  The reversible circuit returned by Algorithm~\ref{alg:hyper} has
  size less than or equal to $(n-1)2^n + 1$ multiple-control Toffoli
  gates.
\end{theorem}

\begin{IEEEproof}
  Notice that each gate application changes at least one bit to its
  correct place. Therefore, Algorithm~\ref{alg:hyper} terminates after
  applying a maximum of $n2^n$ multiple-control Toffoli gates. In
  order to prove an upper bound, we construct a worst-case function
  for this algorithm.

  The first $2^{n-1}$ input patterns match the input, so the most
  significant bit of the output patterns has been completely dealt,
  $2^{n-1}$ zeros are in the upper part of the truth table, the lower
  $2^{n-1}$ must then by definition be $1$. Therefore, starting from
  this step, the most significant bit is fixed. Starting from step
  $2^{n-1}$, flip only the remaining $n-1$ unspecified bits of the
  output.

  Similarly, at step $2^{n-1} + 2^{n-2}$ of the algorithm, the second
  most significant bit will be completely specified. In general, at
  step $2^{n-1} + 2^{n-2} + \ldots + 2^{n-i}$, the $i$ most
  significant bits are completely specified. Thus, the maximum number
  of multiple-control Toffoli gates produced by the
  Algorithm~\ref{alg:hyper} becomes:
  $$n2^{n-1} + (n-1)2^{n-2} + \ldots + n-(n-1)2^{n-n}=(n-1)2^n + 1.$$
  Therefore, $(n-1)2^n + 1$ is an upper bound for the circuit size
  obtained by Algorithm~\ref{alg:hyper}.
\end{IEEEproof}

Using Algorithm~\ref{alg:hyper}, it is possible to find a permutation
for any $n$ that requires at most $(n-1)2^n + 1$ multiple-control
Toffoli gates. If Algorithm~\ref{alg:hyper} runs on right order, then
we find the permutation $[5 \ 2 \ 7 \ 4 \ 1 \ 6 \ 3 \ 0]$ for $n=3$
and the permutation $[5 \ 10 \ 7 \ 4 \ 9 \ 14 \ 11 \ 8 \ 13 \ 2 \
15 \ 12 \ 1 \ 6 \ 3 \ 0]$ for $n=4$. If Algorithm~\ref{alg:hyper} runs
on left order, then we find the permutation $[7 \ 4 \ 1 \ 6 \ 3 \ 0 \
5 \ 2]$ for $n=3$ and the permutation $[15 \ 12 \ 9 \ 14 \ 3 \ 0 \ 13
\ 2 \ 7 \ 4 \ 1 \ 6 \ 11 \ 8 \ 5 \ 10]$ for $n=4$.

Table~\ref{tab:n3} shows the results of applying version of
Algorithm~\ref{alg:hyper} over all $8!=40320$ permutations when $n=3$. We show the number of functions for each gate count and the average number of gates required. 

\section{Analysis of a circuit synthesis based on Cayley graph $I_n$}

In this section, we present the Cayley graph $I_n$ associated to
G-Toffoli gates. These gates are used in the method for circuit synthesis proposed
by Maslov, Dueck and Miller~\cite{mar,mar1}, as we described in
Sec.~\ref{subgeneralized}.

\begin{definition}
  \emph{$C_I$} is the subgroup of $S_{2^n}$, such that all
  permutations $c \in C_I$ are $L_I$-constructible, with only one
  gate.
\end{definition}

\begin{lemma}
  The subgroup $C_I$ is a generating set of $S_{2^n}$.
\end{lemma}

Let $(S_{2^n},\cdot)$ be a finite Symmetric Group with a generating
set $C_I$ given by generalized Toffoli gates.

We denote by \emph{$I_n(V,E)$} the Cayley graph associated with $((S_{2^n},\cdot),C_I)$. The Cayley graph $I_n$ has degree $n2^{n-1}$ and order $2^n!$. Notice that in this case, the corresponding circuits have $n$ gates $N$, $n(n-1)$ gates $C$ and $n(n-1)(n-2)/2$ gates $T$. Also, for $i > 3$ there are $(^{n}_{i})$ generalized Toffoli gates. Therefore,
\[
\sum_{i=1}^{n}i(^{n}_{i})=1(^{n}_{1})+2(^{n}_{2})+3(^{n}_{3})+\ldots+n(^{n}_{n})=n2^{n-1}.
\]

\begin{theorem}
The upper bound for the diameter of the Cayley graph $I_n$ is $(n-1)2^n+1$.
\end{theorem}

\begin{IEEEproof}
Follows directly from Algorithm~\ref{alg:basic}, by construction.
\end{IEEEproof}

\begin{lemma}
  The Cayley graph $I_n$ is not a bipartite graph.
  \label{lem:innb}
\end{lemma}

\begin{IEEEproof}
  We show that the graph $I_n$ has an odd cycle. Let $[3 \ 1 \ 0 \
  2]$, $[3 \ 1 \ 2 \ 0]$, $[1 \ 3 \ 0 \ 2]$, $[0 \ 2 \ 1 \ 3]$ and $[0
  \ 2 \ 3 \ 1]$ be the five vertices of the Cayley graph $I_2$. By
  definition of generating set, we can apply the following changes:
  $(2,3)$, that corresponds to $C^2CNOT(a,b)$ with target in $b$;
  $(0,1)(2,3)$, that corresponds to $C^1CNOT(a)$ with target in $a$;
  $(0,2) (1,3)$, that corresponds to $C^1CNOT(b)$ with target in $b$;
  $(2,3)$, that corresponds to $C^2CNOT(a,b)$ with target in $b$; and
  $(0,2) (1,3)$, that corresponds to $C^1CNOT(b)$ with target in
  $b$. Therefore, we have an odd cycle and the Cayley graph $I_n$ is
  not a bipartite graph.
\end{IEEEproof}

Table~\ref{tab:quantumin} summarizes our analysis of quantum cost for
the synthesis based on Cayley graph $I_n$, showing its relation to an
upper bound for the diameter of the same graph.  The first column
indicates the amount of garbage (ancilla) qubits. The second column
indicates the gate cost, which is upper bound for the diameter of
Cayley graph $I_n$. The third column indicates the quantum cost for
the synthesis based on Cayley graph $I_n$.  This quantum cost is
obtained by multiplying the diameter of the graph by the corresponding
gate count.

\begin{table}
  \begin{center}
    \caption{Analysis of quantum cost for the synthesis based on
      Cayley graph $I_n$.}
    \label{tab:quantumin}
    \begin{tabular}{|c|c|c|}
      \hline
      garbage & gate count (gc) & quantum cost (qc) \\		
      \hline
      0 & $(n-1)2^n + 1$ & $((n-1)2^n + 1)(2^n - 3)$ \\
      1 & $(n-1)2^n + 1$ & $((n-1)2^n + 1)(24n - 88)$ \\
      n-3 & $(n-1)2^n + 1$ & $((n-1)2^n + 1)(10n - 25)$ \\
      \hline
    \end{tabular}
  \end{center}
\end{table}

\section{Analysis of a circuit synthesis based on Cayley graph $H_n$}

In this section, we present the Cayley graph $H_n$ associated to
MC-Toffoli gates. These gates are used in the hypercube method for
circuit synthesis, the we introduced in Sec.~\ref{submultiple}.

\begin{definition}
  \emph{$C_H$} is the subgroup of $S_{2^n}$, such that, all
  permutations $c \in C_H$ are $L_H$-constructible, with only one
  gate.
\end{definition}

\begin{corollary}
  The subgroup $C_H$ is a generating set of~$S_{2^n}$.
\end{corollary}

Let $(S_{2^n},\cdot)$ be a finite Symmetric Group with a generating
set $C_H$ given by multiple-control Toffoli gates.

We denote by \emph{$H_n(V,E)$} the Cayley graph associated with
  $((S_{2^n},\cdot),C_H)$. Notice that the Cayley graph $H_n$ has degree $n2^{n-1}$ and order $2^n!$, the multiple-control Toffoli $C^nNOT(x_1,
x_2,$~$\ldots$~$,x_{n})$ have values $n$ target lines and $0 \leq k
\leq 2^{n-1} - 1$, where $k$ is a decimal value that represent the
lines control. So, we have $n2^{n-1}$ elements in $C_H$.

The Cayley graph $H_n$ has a generating set of the same size and
numbers of vertices of the Cayley graph $I_n$, but those Cayley graphs
are not isomorphic.

\begin{lemma}
  The Cayley graph $H_n$ is a bipartite graph.
  \label{lem:hnb}
\end{lemma}

\begin{IEEEproof}
  Let $x=\{$ $x_1,$ $x_2,$ $\ldots,$ $x_k,$ $x_l,$ $\ldots,$ $x_m$,
  $x_n,$ $\ldots,$ $x_{n2^n-1}$ $\}$ be any vertex of the Cayley graph
  $H_n$ in binary. Let $y=\{$ $y_1,$ $y_2,$ $\ldots,$ $y_l,$ $\ldots,$
  $y_k,$ $\ldots,$ $x_{n2^n-1}$ $\}$ and $z=\{$ $z_1,$ $z_2,$
  $\ldots,$ $z_n,$ $\ldots,$ $z_m,$ $\ldots,$ $x_{n2^n-1}$ $\}$ be
  neighbors of $x$ in binary, where for all $1 \leq i < 2^n$,
  $x_i=y_i$ and $x_i=z_i$, except $i=m$, $i=n$, $i=k$ and $i=l$, in
  which cases $x_m=y_m$, $x_n=y_n$, $x_k=z_k$ and $x_l=z_l$. If $m=k$
  and $n=l$, then $y=z$. If $m=k$ and $n \neq l$, the Hamming distance
  between $y$ and $z$ is $3$, then there is not an edge. If $m \neq k$
  and $n \neq l$, the Hamming distance between $y$ and $z$ is $4$,
  then there is not an edge. Therefore, the Cayley graph $H_n$ does
  not have odd cycles and $H_n$ is bipartite.
\end{IEEEproof}

\begin{theorem}
  The Cayley graph $H_n$ is not isomorphic to the Cayley graph $I_n$.
\end{theorem}

\begin{IEEEproof}
  It follows directly from Lemmas~\ref{lem:innb} and \ref{lem:hnb}.
\end{IEEEproof}

For graphs $H_n$, we have the following results for the upper and
lower bound to the distance of the Cayley graph $H_n$.

\begin{theorem}
  The distance $d(x,\iota)$ between the vertex $x$ of the graph $H_n$
  to the identity element $\iota$ is bounded by $\displaystyle
  d_H(x,\iota)/2 \leq d(x,\iota) < d_H(x,\iota)$.
  \label{theo:dis}
\end{theorem}

\begin{IEEEproof}
  By definition, if we apply a multiple-control Toffoli gate, we may
  have the followings results: i) $2$-move, meaning that two bits
  simultaneouly go to their correct positions; ii) $0$-move, meaning
  that one bit goes to its correct position while another one goes to
  a wrong position; iii) $-2$-move, meaning that two bits go to wrong
  positions.

  The worst case is the reverse permutation. For example, $\rho=[$
  $111,$ $110,$ $\ldots,$ $001,$ $000$ $]$ for $n=3$. In this case,
  the Hamming distance to the identity permutation is
  $n2^n$, i.e., each bit is in the wrong position.
  We need to analyse only two cases, because our algorithm does not apply MC-Toffoli gates on bits already in the correct position, thus avoiding simultaneous misplaces of two bits.
  
  The first case gives us a lower bound, because we have the best case
  if all exchanges are $2$-moves. Therefore, we have
  $d_H(x,\iota)=n2^n/2=n2^{n-1}$. So, $d(x,\iota) \geq
  d_H(x,\iota)/2$.

  The second case gives us an upper bound, because we have the worst
  case if all exchanges are $0$-moves. Therefore, we have
  $d(x,\iota)=n2^n$, but by Algorithm~\ref{alg:hyper} we have an upper
  bound of $(n-1)2^n$. So, $d(x,\iota) < d_H(x,\iota)$.
\end{IEEEproof}

\begin{theorem}
The upper bound for the diameter of the Cayley graph $H_n$ is $(n-1)2^n+1$.
\end{theorem}

\begin{IEEEproof}
Follows directly from Algorithm~\ref{alg:hyper}, by construction.
\end{IEEEproof}

In order to obtain the lower bound for the diameter, we use the fact
that the multiple-control Toffoli gates change the position of two
bits. 

\begin{theorem}
  The lower bound for the diameter of the Cayley graph $H_n$ is
  $n2^{n-1}$.
\end{theorem}

\begin{IEEEproof}
Remark that the reverse permutation has Hamming distance to the identity permutation equal to $n2^n$. The all MC-Toffoli gate applied by Algorithm ~\ref{alg:hyper}, over reverse permutation is a $2$-move. Therefore, we have $n2^{n}/2=n2^{n-1}$.  
\end{IEEEproof}

\begin{table}[ht]
  \begin{center}
    \caption{Analysis of quantum cost on Cayley graph $H_n$}
    \label{tab:quantumhn}
    \begin{tabular}{|c|c|c|}
      \hline
      garbage & gate count (gc) & quantum cost (qc) \\		
      \hline
      0 & $(n-1)2^n + 1$ & $((n-1)2^n + 1)(2^n - 3 + 2m)$ \\
      1 & $(n-1)2^n + 1$ & $((n-1)2^n + 1)(24n - 86)$ \\
      n-3 & $(n-1)2^n + 1$ & $((n-1)2^n + 1)(10n - 23)$ \\
      \hline
    \end{tabular}
  \end{center}
\end{table}

Table~\ref{tab:quantumhn} summarizes our analysis of quantum cost for
the synthesis based on Cayley graph $H_n$, showing its relation to an
upper bound for the diameter of the same graph.  The first column
indicates the amount of garbage (ancilla) qubits. The second column
indicates the gate count, which is an upper bound for the diameter of
Cayley graph $H_n$. The third column indicates the quantum cost for
the synthesis based on Cayley graph $H_n$.  This quantum cost is
obtained by multiplying the diameter of the graph by the corresponding
gate count.

\section{Conclusions}

Since reversibility is an essential aspect the circuit model of quantum computers, we
must have efficient methods for designing and analysing reversible
circuits.  Group Theory provides a unified framework for the
development and analysis of methods for reversible circuit
synthesis. In this work, we studied two Cayley graphs, $I_n$ and
$H_n$, that can be applied to the synthesis of reversible circuits.

Maslov, Dueck and Miller~\cite{mar,mar1} proposed an algorithm to
reversible circuit synthesis using generalized Toffoli gates that we
modeled by the Cayley graph $I_n$. The framework of the theory of
Cayley graphs enabled us to prove that the diameter of $I_n$ is less
than $(n-1)2^n+1$ and the number of vertices is $2^n!$.  These bounds
are consistent with the gate count and quantum cost complexity of the
circuit synthesis using G-Toffoli gates.

We presented an algorithm to reversible circuit synthesis using
multiple-control Toffoli gates. The proposed circuit synthesis is
based on the hypercube Cayley graph $H_n$. The diameter of the Cayley
graph $H_n$ is at most $(n-1)2^n+1$ and is at least $n2^{n-1}$. Since
the number of vertices of $H_n$ is $2^n!$, we have that the number of
vertices is a factorial on the diameter. We proved that Cayley graph
$H_n$ is not isomorphic to Cayley graph $I_n$, so the corresponding
synthesis algorithms are different.  These bounds are consistent with
the gate count and quantum cost complexity of the circuit synthesis
using MC-Toffoli gates.

We may expect Cayley graphs to be an attractive and versatile
framework for analysing reversible circuit synthesis.

\section*{Acknowledgements}

The authors thank Luís Cunha and Roberto Sampaio for helpful
discussions.

\bibliographystyle{IEEEtran} 
\bibliography{weciq2012BibTex}

\begin{thebibliography}{10}
\providecommand{\url}[1]{#1}
\csname url@samestyle\endcsname
\providecommand{\newblock}{\relax}
\providecommand{\bibinfo}[2]{#2}
\providecommand{\BIBentrySTDinterwordspacing}{\spaceskip=0pt\relax}
\providecommand{\BIBentryALTinterwordstretchfactor}{4}
\providecommand{\BIBentryALTinterwordspacing}{\spaceskip=\fontdimen2\font plus
\BIBentryALTinterwordstretchfactor\fontdimen3\font minus
  \fontdimen4\font\relax}
\providecommand{\BIBforeignlanguage}[2]{{%
\expandafter\ifx\csname l@#1\endcsname\relax
\typeout{** WARNING: IEEEtran.bst: No hyphenation pattern has been}%
\typeout{** loaded for the language `#1'. Using the pattern for}%
\typeout{** the default language instead.}%
\else
\language=\csname l@#1\endcsname
\fi
#2}}
\providecommand{\BIBdecl}{\relax}
\BIBdecl

\bibitem{nielsen00}
M.~A. Nielsen and I.~L. Chuang, \emph{Quantum Computation and Quantum
  Information}.\hskip 1em plus 0.5em minus 0.4em\relax Cambridge University
  Press, 2000.

\bibitem{toffoli}
T.~Toffoli, ``Reversible computing,'' in \emph{Proceedings of the 7th
  Colloquium on Automata, Languages and Programming}.\hskip 1em plus 0.5em
  minus 0.4em\relax London, UK, UK: Springer-Verlag, 1980, pp. 632--644.

\bibitem{rolf}
R.~Landauer, ``{Irreversibility and heat generation in the computing
  process},'' \emph{IBM Journal of Research and Development}, vol.~5, pp.
  183--191, 1961.

\bibitem{ben}
C.~H. Bennett, ``Logical reversibility of computation,'' \emph{IBM J. Res.
  Dev.}, vol.~17, no.~6, pp. 525--532, 1973.

\bibitem{survey}
\BIBentryALTinterwordspacing
M.~Saeedi and I.~L. Markov, ``Synthesis and optimization of reversible circuits
  - a survey,'' \emph{CoRR}, vol. abs/1110.2574, 2011. [Online]. Available:
  \url{http://arxiv.org/abs/1110.2574}
\BIBentrySTDinterwordspacing

\bibitem{leo}
L.~Storme, A.~D. Vos, and G.~Jacobs, ``Group theoretical aspects of reversible
  logic gates,'' \emph{Journal of Universal Computer Science}, vol.~5, no.~5,
  pp. 307--321, jan 1999.

\bibitem{alexis}
A.~Devos, B.~Raa, and L.~Storme, ``{Generating the group of reversible logic
  gates},'' \emph{Journal of Physics A Mathematical General}, vol.~35, pp.
  7063--7078, Aug. 2002.

\bibitem{mar}
D.~Maslov, G.~W. Dueck, and D.~M. Miller, ``Toffoli network synthesis with
  templates,'' \emph{IEEE Trans. on CAD of Integrated Circuits and Systems},
  vol.~24, p. 2005, 2005.

\bibitem{mar1}
D.~M. Miller, D.~Maslov, and G.~W. Dueck, ``A transformation based algorithm
  for reversible logic synthesis,'' in \emph{Proceedings of the 40th annual
  Design Automation Conference}, ser. DAC '03.\hskip 1em plus 0.5em minus
  0.4em\relax New York, NY, USA: ACM, 2003, pp. 318--323.

\bibitem{vadapalli}
P.~Vadapalli and P.~K. Srimani, ``A new family of cayley graph interconnection
  networks of constant degree four,'' \emph{IEEE Trans. Parallel Distrib.
  Syst.}, vol.~7, no.~1, pp. 26--32, Jan. 1996.

\bibitem{she}
V.~V. Shende, A.~K. Prasad, I.~L. Markov, and J.~P. Hayes, ``Synthesis of
  reversible logic circuits,'' \emph{IEEE Trans. on CAD of Integrated Circuits
  and Systems}, vol.~22, no.~6, pp. 710--722, 2003.

\bibitem{Barenco1995}
A.~Barenco, C.~H. Bennett, R.~Cleve, D.~P. DiVincenzo, N.~Margolus, P.~Shor,
  T.~Sleator, J.~A. Smolin, and H.~Weinfurter, ``{Elementary gates for quantum
  computation},'' \emph{Physical Review A}, vol.~52, no.~5, pp. 3457--3467,
  Nov. 1995.

\bibitem{mar2}
D.~Maslov, C.~Young, G.~W. Dueck, and D.~M. Miller, ``Quantum circuit
  simplification using templates,'' \emph{Design, Automation and Test in
  Europe}, pp. 1208--1213, Mar. 2005.

\bibitem{wille1}
\BIBentryALTinterwordspacing
R.~Wille, M.~Saeedi, and R.~Drechsler, ``Synthesis of reversible functions
  beyond gate count and quantum cost,'' 2010. [Online]. Available:
  \url{http://arxiv.org/abs/1004.4609}
\BIBentrySTDinterwordspacing

\bibitem{luis}
L.~A.~B. Kowada, ``Construção de algoritmos reversíveis e quânticos,'' Ph.D.
  dissertation, Universidade Federal do Rio de Janeiro, Programa de Engenharia
  de Sistemas e Computação, Rio de Janeiro, 2006.

\bibitem{dmi}
\BIBentryALTinterwordspacing
D.~Maslov and M.~Saeedi, ``Reversible circuit optimization via leaving the
  boolean domain,'' \emph{CoRR}, vol. abs/1103.0215, 2011. [Online]. Available:
  \url{http://arxiv.org/abs/1103.0215}
\BIBentrySTDinterwordspacing

\end{thebibliography}

\end{document}